\documentclass[pre,aps,amsfonts,amssymb,showpacs,preprint]{revtex4}

\usepackage{graphicx,bm}
\def\x{{\bm x}}
\def\k{{\bm k}}
\def\u{{\bm u}}

\def\el{{\bm \ell}}
\def\m{{\bm m}}
\def\n{{\bm n}}
\def\bomega{{\bm\omega}}

\def\ltw{{\hbox{${\lower 6pt\hbox{$<$}}\atop{\raise 2pt\hbox{$\sim$}}$}\,}}
\def\gtw{{\hbox{${\lower 6pt\hbox{$>$}}\atop{\raise 2pt\hbox{$\sim$}}$}\,}}

\begin{document}

\title{Effective degrees of nonlinearity in a family of generalized 
models of two-dimensional turbulence}

\author{Chuong V. Tran, David G. Dritschel, and Richard K. Scott}

\affiliation{School of Mathematics and Statistics, University of St Andrews,
St Andrews KY16 9SS, United Kingdom}

\date{\today}

\begin{abstract}

We study the small-scale behavior of generalized two-dimensional 
turbulence governed by a family of model equations, in which the 
active scalar $\theta=(-\Delta)^{\alpha/2}\psi$ is advected by the 
incompressible flow $\u=(-\psi_y,\psi_x)$. Here $\psi$ is the stream 
function, $\Delta$ is the Laplace operator and $\alpha$ is a positive 
number. The dynamics of this family are characterized by the material 
conservation of $\theta$, whose variance $\langle\theta^2\rangle$ is 
preferentially transferred to high wave numbers (direct transfer). As 
this transfer proceeds to ever-smaller scales, the gradient 
$\nabla\theta$ grows without bound. This growth is due to the stretching
term $(\nabla\theta\cdot\nabla)\u$ whose ``effective degree of nonlinearity'' 
differs from one member of the family to another. This degree depends
on the relation between the advecting flow $\u$ and the active scalar 
$\theta$ (i.e. on $\alpha$) and is wide ranging, from approximately linear 
to highly superlinear. Linear dynamics are realized when $\nabla\u$ is 
a quantity of no smaller scales than $\theta$, so that it is insensitive 
to the direct transfer of the variance of $\theta$, which is nearly 
passively advected. This case corresponds to $\alpha\ge2$, for which 
the growth of $\nabla\theta$ is approximately exponential in time and 
non-accelerated. For $\alpha<2$, superlinear dynamics are realized as 
the direct transfer of $\langle\theta^2\rangle$ entails a growth in 
$\nabla\u$, thereby enhancing the production of $\nabla\theta$. This 
superlinearity reaches the familiar quadratic nonlinearity of 
three-dimensional turbulence at $\alpha=1$ and surpasses that for 
$\alpha<1$. The usual vorticity equation ($\alpha=2$) is the border 
line, where $\nabla\u$ and $\theta$ are of the same scale, separating 
the linear and nonlinear regimes of the small-scale dynamics. We discuss 
these regimes in detail, with an emphasis on the locality of the direct 
transfer.

\end{abstract}


\maketitle

\section{INTRODUCTION}

The production of progressively smaller scales, possibly to be limited 
by viscous effects only, in incompressible fluid flow at high Reynolds 
numbers is a fundamental problem in fluid dynamics. This long-standing
problem is of genuine interest for obvious reasons. One is that the 
production of small scales plays a key role in the possible development 
of singularities from smooth initial conditions in the three-dimensional 
(3D) Euler or Navier--Stokes equations that govern the flow. Another 
reason is that in the presence of a large-scale forcing, a persistent 
production of small scales would be crucial to maintain a spectral energy 
flux (direct energy cascade). The realizability of such a steady and 
viscosity independent flux is central to the Kolmogorov theory of 
turbulence as this would be required to rid the virtually inviscid energy 
inertial range of the injected energy, thereby making it possible for a 
statistical equilibrium to be established. This dynamical scenario is 
either explicitly or implicitly assumed to apply to other fluid systems 
as well, not just the 3D Navier--Stokes equations. For example, in the 
Kraichnan--Batchelor \cite{Kraichnan67,Kraichnan71,Batchelor69} theory 
of two-dimensional (2D) turbulence, the dynamics of the mean-square 
vorticity (twice the enstrophy) are assumed to be synonymous in many 
aspects to those of the 3D energy. In particular, the enstrophy injected 
into the system at large scales is hypothesized to cascade to a dissipation 
range at small scales. As another example, the mean-square potential 
vorticity in the quasi-geostrophic geophysical flow model is believed to 
behave in a similar manner \cite{Charney71}. Thus ``cascading dynamics'' 
have been considered universal among fluid systems. 

The evolution of fluid flow is intrinsically nonlinear because of the 
quadratic advection term, which couples all scales of motion. Apparently, 
this is an underpinning reason for the cascade universality mentioned in 
the preceding paragraph. However, the ``effective degree of nonlinearity'' 
of the small-scale dynamics is not always quadratic and differs from 
one system to another. The implication is that the presumed cascades
would have fundamental differences and would not be universal in a 
strict sense. For an example of the discrepancy in the effective degree 
of nonlinearity among fluid systems, let us consider the respective 
evolution equations for the 3D vorticity $\bomega$ and 2D vorticity 
gradient $\nabla\omega$ given by
\begin{eqnarray}
\label{3dvorticity}
\partial_t\bomega + (\u\cdot\nabla)\bomega = (\bomega\cdot\nabla)\u,
~~~\nabla\cdot\u=0
\end{eqnarray}
and
\begin{eqnarray}
\label{2dgradient}
\partial_t\nabla\omega + (\u\cdot\nabla)\nabla\omega = 
\omega\n\times\nabla\omega-(\nabla\omega\cdot\nabla)\u,~~~\nabla\cdot\u=0,
\end{eqnarray}
where $\u$ is the fluid velocity and $\n$ is the normal to the fluid 
domain in 2D. The stretching term $(\bomega\cdot\nabla)\u$ for the 3D 
vorticity $\bomega$ in Eq.\ (\ref{3dvorticity}) is essentially quadratic 
in $\bomega$ because the velocity gradient $\nabla\u$ is expected to
behave as $\bomega$ on phenomenological grounds. As a consequence, an 
explosive 3D vorticity growth from a smooth initial vorticity field is 
possible, if not inevitable \cite{Kerr93,Kerr05}. In contrast, the 
stretching term $(\nabla\omega\cdot\nabla)\u$ for the 2D vorticity 
gradient $\nabla\omega$ in Eq.\ (\ref{2dgradient}) is virtually linear 
in $\nabla\omega$ because $\nabla\u$ is well behaved in the sense that 
the mean square vorticity 
$\langle\omega^2\rangle=\langle|\nabla\u|^2\rangle$ is conserved. 
(Note that the rotation term $\omega\n\times\nabla\omega$ does not 
affect the amplitude of $\nabla\omega$.) As a result, the growth of 
2D vorticity gradients can possibly be approximately exponential in 
time only, a relatively mild behavior. Hence, one would expect profound 
differences between the (highly nonlinear) 3D vorticity and the (nearly 
linear) 2D vorticity gradient dynamics. A notable example of these
differences is that in the inviscid limit, the 2D enstrophy dissipation 
rate vanishes \cite{Dmitruk05,TD06a,DTS07}, whereas the 3D energy 
dissipation rate presumably remains nonzero. Another example is the 
discrepancy in the dependence on the Reynolds number of the number 
of degrees of freedom in the two cases \cite{T09}.

The effective degree of nonlinearity in the above sense differs not only 
between 2D and 3D fluids but also among 2D fluid systems. In this study 
we investigate this varying degree among members of a broad family of 
generalized models of 2D turbulence, first introduced by Pierrehumbert, 
Held, and Swanson \cite{Pierrehumbert94}. By so doing we extend several 
previous studies \cite{Schorghofer00,Smith02,Wantanabe04,T04}, aiming to 
unify our understanding of turbulent transfer in physically realizable 
fluid systems. The family's dynamics are characterized by the material 
conservation of the active scalar $\theta=(-\Delta)^{\alpha/2}\psi$, whose 
variance $\langle\theta^2\rangle$ is preferentially transferred to high
wave numbers (small scales).  Here $\psi$ is the stream function,
$\Delta$ is Laplace's operator, and $\alpha$ is a positive number. As
the transfer of $\langle\theta^2\rangle$ proceeds to ever-smaller
scales, the gradient $\nabla\theta$ grows without bound. This growth
is due to the stretching term $(\nabla\theta\cdot\nabla)\u$ whose
effective degree of nonlinearity depends on $\alpha$ and is wide
ranging, from approximately linear to highly superlinear.  Linear 
behavior is realized when $\nabla\u$ is a quantity of no smaller scales 
than $\theta$, so that the transfer of $\langle\theta^2\rangle$ to the 
small scales (direct transfer) has no significant effects on $\nabla\u$.  
In other words, $\theta$ behaves nearly passively.  This case corresponds 
to $\alpha\ge2$, for which $\nabla\theta$ can grow approximately
exponentially in time without acceleration. For $\alpha<2$,
superlinear dynamics can be realized as the direct transfer of
$\langle\theta^2\rangle$ entails a growth in $\nabla\u$, thereby 
enhancing the production of $\nabla\theta$. This superlinearity reaches 
the familiar quadratic nonlinearity of three-dimensional turbulence at 
$\alpha=1$ and exceeds that for $\alpha<1$. The usual vorticity equation
($\alpha=2$) is the border line, where $\nabla\u$ and $\theta$ are of
the same scale ($\langle|\nabla\u|^2\rangle=\langle\theta^2\rangle$),
separating the linear and nonlinear regimes of the small-scale dynamics. 
We discuss these dynamical regimes in detail, with an emphasis on the 
local nature of the transfer of $\langle\theta^2\rangle$. The
implication of the present results is that a comprehensive theory for
this family of generalized 2D turbulence needs to account for the wide
range of effective degrees of nonlinearity of the family's small-scale
dynamics.

\section{Governing equations}

The equation governing the evolution of the family of active scalars 
$\theta=(-\Delta)^{\alpha/2}\psi$ (for $\alpha>0$) advected by the 
incompressible flow $\u=(-\psi_y,\psi_x)$ is 
\begin{eqnarray}
\label{governing}
\theta_t + \u\cdot\nabla\theta &=& 0.
\end{eqnarray}
This equation has been proposed by Pierrehumbert, Held, and Swanson
\cite{Pierrehumbert94} in an attempt to better understand the nature 
of transfer locality in 2D turbulence, by examining how turbulent
transfer responses to changes in the parameter $\alpha$. 
Equation (\ref{governing}) is 
physically relevant for selected values of $\alpha$. The usual 2D 
vorticity equation corresponds to $\alpha=2$. When $\alpha=1$, Eq.\ 
(\ref{governing}) is known as the surface quasi-geostrophic equation 
and governs the advection of the potential temperature, which is 
proportional to $\theta=(-\Delta)^{1/2}\psi$, on the surface of a 
quasi-geostrophic fluid. In addition to the genuine interest due to 
this physical significance \cite{Held95,Schorghofer00,Smith02, 
Wantanabe04,TB05,Scott06,T04,T06,TD06b}, the surface quasi-geostrophic 
equation has received some special attention for its resemblance to 
the 3D Euler system \cite{Constantin94a,Constantin94b,Ohkitani97, 
Cordoba04}. A mathematical feature of particular interest is the 
possible development of finite-time singularities (from smooth initial 
conditions), which, as argued by pioneering studies \cite{Bennett84,
Constantin94a,Constantin94b} of this problem, could be associated with 
the formation of weather fronts in the atmosphere. This, however, appears
not to be the case \cite{Cordoba98}.

For simplicity, we consider Eq.\ (\ref{governing}) in a doubly periodic 
domain of size $L$, and all fields concerned are assumed to have zero 
spatial average. This allows us to express the stream function as
\begin{eqnarray}
\label{Fourier}
\psi(\x,t) &=& \sum_{\k}\widehat\psi(\k,t)\exp\{i\k\cdot\x\}.
\end{eqnarray}
Here $\k=2\pi L^{-1}(k_x,k_y)$, where $k_x$ and $k_y$ are integers 
not simultaneously zero. The reality of $\psi$ requires
$\widehat\psi(\k,t)=\widehat\psi^*(-\k,t)$, where the asterisk denotes 
the complex conjugate. The fractional derivative $(-\Delta)^{\alpha/2}$ 
(which can be readily extended to $\alpha<0$, though not considered in 
this study) is defined by
\begin{eqnarray}
\label{Fourier1}
\theta(\x,t) = (-\Delta)^{\alpha/2}\psi(\x,t) &=& \sum_{\k} k^{\alpha}\,
\widehat\psi(\k,t)\exp\{i\k\cdot\x\}\nonumber\\
&=& \sum_{\k}\widehat\theta(\k,t)\exp\{i\k\cdot\x\},
\end{eqnarray}
where $k=|\k|$ is the wave number. Equation (\ref{governing}) expresses 
material conservation of $\theta$, which gives rise to an infinite set
of conserved quantities. In particular, the generalized enstrophy
(active scalar variance)
\begin{eqnarray}
Z &=& \frac{1}{2}\langle\theta^2\rangle =
\frac{1}{2}\langle|(-\Delta)^{\alpha/2}\psi|^2\rangle = 
\frac{1}{2}\sum_{\k} k^{2\alpha}\,|\widehat\psi(\k,t)|^2 
\end{eqnarray}
is conserved. In addition, the generalized energy
\begin{eqnarray}
E &=& \frac{1}{2}\langle\psi\theta\rangle =
\frac{1}{2}\langle|(-\Delta)^{\alpha/4}\psi|^2\rangle = 
\frac{1}{2}\sum_{\k} k^{\alpha}\,|\widehat\psi(\k,t)|^2
\end{eqnarray}
is also conserved. Note that $E$ is the usual kinetic energy when 
$\alpha=2$, while $Z$ is the usual kinetic energy when $\alpha=1$.
Only for these cases is the kinetic energy conserved. The modal powers 
(spectra) of $E$ and $Z$ differ by the factor $k^\alpha$. Therefore, the 
redistribution of a non-negligible amount of $E$ to small scales 
would violate the conservation of $Z$. Similarly, the redistribution of 
a non-negligible amount of $Z$ to large scales would violate the 
conservation of $E$. This means that if a spectrally localized profile 
is to spread out in wave number space, most of $E$ and $Z$ get transferred 
to large and small scales, respectively. This is the basis for the 
dual cascade hypothesis in 2D turbulence. Here we are mainly concerned 
with the direct transfer of $Z$. A more complete treatment should include 
the inverse transfer of $E$ as well since these are known to be intimately 
related. 

Given Eq.\ (\ref{Fourier}), we can express $\u=(-\psi_y,\psi_x)$ in terms 
of a Fourier series in the form
\begin{eqnarray}
\label{Fourier2}
\u(\x,t) &=& i\sum_{\k}(-k_y,k_x)\widehat\psi(\k,t)\exp\{i\k\cdot\x\}.
\end{eqnarray}
By substituting Eqs.\ (\ref{Fourier1}) and (\ref{Fourier2}) into Eq.\ 
(\ref{governing}) we obtain the evolution equation for each individual 
Fourier mode $\widehat\theta(\k,t)=k^\alpha\,\widehat\psi(\k,t)$ of 
the conserved quantity $\theta$
\begin{eqnarray}
\label{mode-evol}
\frac{d}{dt}\widehat\theta(\k,t) &=& \sum_{\el+\m=\k}
\frac{(m^\alpha-\ell^\alpha)\,\el\times\m}{\ell^\alpha m^\alpha}\,
\widehat\theta(\el,t)\widehat\theta(\m,t),
\end{eqnarray}
where $\el\times\m=\ell_xm_y-\ell_ym_x$. The sum on the right-hand side 
of Eq.\ (\ref{mode-evol}) involves all modes (except $\widehat\theta(\k,t)$) 
and is a measure of the level 
of `excitation' of the mode $\widehat\theta(\k,t)$ due to all admissible 
wave vector triads $\k=\el+\m$. For a given triad, the coupling coefficient 
$(m^\alpha-\ell^\alpha)\,\el\times\m/(\ell^\alpha m^\alpha)$ depends 
on $\alpha$. Its magnitude, together with the magnitudes of
the coupling coefficients in the governing equations for
$\widehat\theta(\el,t)$ and $\widehat\theta(\m,t)$, is a measure of triad 
dynamical activity, in the sense that larger (in magnitude) coupling 
coefficients correspond to more intense modal dynamics. This is intimately 
related to the effective degree of nonlinearity and locality of the 
small-scale dynamics as will be seen in the subsequent sections.

\section{Effective degrees of nonlinearity of the small-scale dynamics}

We now examine the behavior of $\nabla\theta$. Generally speaking, any 
derivative $(-\Delta)^{\eta}\theta$, for $\eta>0$, can be 
called a small-scale quantity. Here we consider $\nabla\theta$, which 
is a ``twin brother'' of $(-\Delta)^{1/2}\theta$, for its special 
status in Eq.\ (\ref{governing}) as well as its mathematical tractability. 
For $\alpha=2$, a similar treatment of $\Delta\theta=-\Delta\omega$ can 
be carried out in the same manner.

\subsection{Growth of the active scalar gradient}
 
The governing equation for $\nabla\theta$ is
\begin{eqnarray}
\label{Governing}
\partial_t \nabla\theta + (\u\cdot\nabla)\nabla\theta =
\nabla\times\u\times\nabla\theta - (\nabla\theta\cdot\nabla)\u, 
\end{eqnarray}
which can be obtained by replacing $\omega$ in Eq.\ (\ref{2dgradient}) 
by $\theta$. Like Eq.\ (\ref{2dgradient}), the effect of the first term 
on the right-hand side of Eq.\ (\ref{Governing}) is to rotate 
$\nabla\theta$ without changing its magnitude. The amplification of 
$\nabla\theta$ is due solely to the stretching term 
$(\nabla\theta\cdot\nabla)\u$ and is governed by
\begin{eqnarray}
\label{Governing1}
\partial_t |\nabla\theta| + (\u\cdot\nabla)|\nabla\theta| =
- \frac{\nabla\theta}{|\nabla\theta|}\cdot(\nabla\theta\cdot\nabla)\u 
\le |\nabla\u||\nabla\theta|.
\end{eqnarray}
Equation (\ref{Governing1}) implies that following the fluid motion, 
$|\nabla\theta|$ can grow exponentially in time with an instantaneous 
rate bounded from above by $|\nabla\u|$. Hence the behavior of 
$|\nabla\u|$ holds the key to understanding the dynamics of $\nabla\theta$. 
Evidently, following the trajectory of a fluid 
``particle'' starting from $\x=\x_0$ at $t=0$, the growth of 
$|\nabla\theta|$ is formally constrained by 
\begin{eqnarray}
|\nabla\theta| &\le& 
|\nabla\theta_0|\exp\left\{\int_0^t|\nabla\u|\,d\tau\right\},
\end{eqnarray}
where $\theta_0=\theta(\x_0,0)$ and the integral is along the 
trajectory in question. Hence, on average, the rate $r$ defined by
\begin{eqnarray}
\label{ratebound}
r &=& \frac{1}{t}\int_0^t|\nabla\u|\,d\tau
\end{eqnarray}
provides an upper bound for the exponential growth rate of
$|\nabla\theta|$. Note that for $\alpha=1$
($\langle|\nabla\u|^2\rangle=\langle|\nabla\theta|^2\rangle$),
a double exponential growth of $|\nabla\theta|$ is allowed but not 
necessarily implied by the preceding equations. Nevertheless, it is 
interesting to note that Ohkitani and Yamada \cite{Ohkitani97} 
observed such a behavior in their simulations, thereby suggesting 
a negative answer to the question of finite-time singularities in 
the surface quasi-geostrophic equation. This is consistent with
the proof of nonexistence of blowup by C\'ordoba \cite{Cordoba98}. 

\subsection{Linear versus nonlinear growth of $\nabla\theta$}

Now for a sense of the behavior of $r$, we consider 
$\langle|\nabla\u|^2\rangle^{1/2}$, which bounds $\langle|\nabla\u|\rangle$
from above by the Cauchy--Schwarz inequality 
$\langle|\nabla\u|\rangle\le\langle|\nabla\u|^2\rangle^{1/2}$. 
For $\alpha\in[2,4]$, $\langle|\nabla\u|^2\rangle^{1/2}$
can be estimated in terms of the inviscid invariants using the following 
version of the H\"older inequality (see, for example, \S\,5 of Ref.\ 
\cite{T04})
\begin{eqnarray}
\label{inter}
\langle|\nabla\u|^2\rangle^{1/2} &\le& 
\langle|(-\Delta)^{\alpha/4}\psi|^2\rangle^{1-2/\alpha}
\langle|(-\Delta)^{\alpha/2}\psi|^2\rangle^{2/\alpha-1/2} \nonumber\\
&=& 
E^{1-2/\alpha}Z^{2/\alpha-1/2}.
\end{eqnarray}
So $\langle|\nabla\u|^2\rangle^{1/2}$ is controlled by the inviscid
invariants $E$ and $Z$. For $\alpha\notin[2,4]$, inequality (\ref{inter}) 
reverses direction. Furthermore, if an initial distribution of $\theta$ 
is to forever spread out in wave number space, 
$\langle|\nabla\u|^2\rangle^{1/2}$ increases without bound for this 
case. This implies that there exist different regimes of $\alpha$
for which $\nabla\u$ evolves quite differently, and the active scalar
gradient dynamics can be characteristically distinct. We discuss all 
these regimes in what follows.

For $\alpha<2$, the divergence of $\langle|\nabla\u|^2\rangle^{1/2}$ 
entails an accelerated growth of $\nabla\theta$ from an exponential 
one. This is the superlinear regime discussed in the introductory 
section. This superlinearity reaches the usual quadratic nonlinearity 
of 3D turbulence at $\alpha=1$, where 
$\langle|\nabla\u|^2\rangle=\langle|\nabla\theta|^2\rangle$. Hence, the
surface quasi-geostrophic and 3D Euler equations are analogous in this
aspect. However, the analogy appears to be superficial as the surface 
quasi-geostrophic equation turns out to be far more ``manageable'' 
than its 3D counterpart---a consequence of the material conservation of
$\theta$. For example, a number of global regularity results have been 
proved for the surface quasi-geostrophic equation, by making use of mild 
dissipation mechanisms represented by $(-\Delta)^{\eta}$ with $\eta\ge1/2$ 
\cite{Carrillo08,Constantin08,Ju05,Kiselev07}, which can be much weaker 
than the usual viscosity. Whereas 
for the 3D Navier--Stokes system, viscosity appears to be inadequate for 
the same purpose. For $\alpha<1$, this quadratic nonlinearity is surpassed 
as the ratio $\langle|\nabla\u|^2\rangle/\langle|\nabla\theta|^2\rangle$ 
diverges in the limit $\langle|\nabla\theta|^2\rangle\to\infty$ because 
\begin{eqnarray}
\langle|\nabla\theta|^2\rangle^{2-\alpha} &\le& 
\langle|\nabla\u|^2\rangle\langle\theta^2\rangle^{1-\alpha}
\end{eqnarray}
(cf.\ Ref.\ \cite{T04}). Active scalar gradient production can then become 
highly intense.

For $\alpha\in[2,4]$, $\nabla\u$ is well behaved in the sense that its
mean square is bounded from above in terms of the inviscid invariants,
[see Eq.\ (\ref{inter})]. In this case, $\nabla\u$ is virtually unaffected 
by the direct transfer of $\langle\theta^2\rangle$. At large $t$, a general 
fluid trajectory is likely to have traversed the domain many times. The 
time average in Eq.\ (\ref{ratebound}) may therefore be approximately 
replaced by the spatial average. Hence, we can write 
\begin{eqnarray}
r \approx \langle|\nabla\u|\rangle \le \langle|\nabla\u|^2\rangle^{1/2}
\le E^{1-2/\alpha}Z^{2/\alpha-1/2},
\end{eqnarray}  
where we have used the Cauchy--Schwarz inequality and Eq.\ (\ref{inter}).
This approximation of $r$ means that $\nabla\theta$ can grow exponentially
in time without acceleration. Thus, approximately linear small-scale 
dynamics can be expected. Note that $\theta$ behaves almost as a passive 
scalar in this regime. The analogy between this case and that of a passive 
scalar was suggested by Schorghofer \cite{Schorghofer00} on 
phenomenological grounds.

When $\alpha>4$, inequality (\ref{inter}) reverses direction, and
$\langle|\nabla\u|^2\rangle^{1/2}$ can no longer be controlled by
the inviscid invariants. However, unlike the case $\alpha<2$, for
which $\langle|\nabla\u|^2\rangle^{1/2}$ diverges toward small scales,
when $\alpha>4$ velocity gradients can be produced at increasingly
large scales only. This production depends on the inverse transfer 
of the generalized energy $E$ (Tran 2004). Within the direct transfer 
range, i.e. the generalized enstrophy range, the portion of 
$\langle|\nabla\u|^2\rangle$, say $\Omega$, cannot increase and instead
remains bounded from above in terms of $Z$. More precisely, as the 
spectra of $\langle|\nabla\u|^2\rangle$ and $Z$ differ by the factor 
$k^{2\alpha-4}$, we have $\Omega\le 2k_*^{4-2\alpha}Z$ (Poincar\'e 
type inequality), where $k_*$ is the lower wave number end of the 
generalized enstrophy range. This suggests that no significant changes 
in the effective degree of nonlinearity of the small-scale dynamics 
occur when $\alpha$ exceeds $4$. Thus, we can expect approximately 
linear small-scale behavior for all $\alpha\ge2$.

In passing, it is worth mentioning that while the small-scale dynamics 
appear to be insensitive to $\alpha$ in the regime $\alpha>2$, the 
large-scale dynamics can vary dramatically. The reason is that for 
large $\alpha$, $\u$ is prone to divergence toward large scales as the 
inverse transfer of $E$ proceeds. This undoubtedly intensifies motions 
at large scales. One may adapt the present notion of degree of 
nonlinearity for a quantitative measure of the large-scale dynamics. 
Analogous to the traditional problem of regularity, which is concerned 
with the possible divergence of $\nabla\theta$, there is a potential 
problem that $\u$ becomes divergent for sufficiently large $\alpha$ if 
the fluid is unbounded. This interesting problem is left for a future study.

\section{Locality of the small-scale dynamics}

This section is concerned with the small-scale dynamics at the modal
level. We establish a connection between the degree of nonlinearity 
and dynamical activity of typical local triads at small scales. 
Here the dynamical activity of a given triad is associated with the 
magnitude of the coupling coefficients within the triad and is 
independent of the amplitude of the three modal members. These local
triads are shown to be highly active for $\alpha<2$ and moderately active 
for $\alpha=2$, but become virtually inactive for $\alpha>2$. This implies 
that higher effective degrees of nonlinearity correspond to more dynamically 
intense local triads. Thus, the effective degree of nonlinearity is also a 
measure of dynamical activity of local triads at small scales. The 
transition at $\alpha=2$ from high activity to virtually no activity
of local triads is consistent with phenomenological arguments 
\cite{Pierrehumbert94} that the generalized enstrophy cascade is spectrally 
local for $\alpha<2$, but becomes dominated by nonlocal interactions for 
$\alpha>2$. Below, we also examine the dynamics of nonlocal triads and 
elaborate on the nature of the locality transition, in order to provide
a detailed picture of the direct transfer of $\langle\theta^2\rangle$ 
at the modal level.

Within each individual triad $\k=\el+\m$, the transfer of modal
generalized enstrophy is governed by
\begin{eqnarray}
\label{triad}
\frac{d}{dt}|\widehat\theta(\k)|^2 &=& 
\frac{(m^\alpha-\ell^\alpha)\,\el\times\m}{m^\alpha \ell^\alpha}
\left(\widehat\theta(\el)\widehat\theta(\m)\widehat\theta^*(\k)
+ \widehat\theta^*(\el)\widehat\theta^*(\m)\widehat\theta(\k)\right)
\nonumber\\
&=& 
C_\k\left(\widehat\theta(\el)\widehat\theta(\m)\widehat\theta^*(\k)
+ \widehat\theta^*(\el)\widehat\theta^*(\m)\widehat\theta(\k)\right),
\nonumber\\
\frac{d}{dt}|\widehat\theta(\el)|^2 &=& 
\frac{(k^\alpha-m^\alpha)\,\el\times\m}{k^\alpha m^\alpha}
\left(\widehat\theta(\k)\widehat\theta^*(\m)\widehat\theta^*(\el)
+ \widehat\theta^*(\k)\widehat\theta(\m)\widehat\theta(\el)\right)
\nonumber\\
&=&
C_\el\left(\widehat\theta(\k)\widehat\theta^*(\m)\widehat\theta^*(\el)
+ \widehat\theta^*(\k)\widehat\theta(\m)\widehat\theta(\el)\right),
\nonumber\\
\frac{d}{dt}|\widehat\theta(\m)|^2 &=& 
\frac{(\ell^\alpha-k^\alpha)\,\el\times\m}{\ell^\alpha k^\alpha}
\left(\widehat\theta(\k)\widehat\theta^*(\el)\widehat\theta^*(\m)
+ \widehat\theta^*(\k)\widehat\theta(\el)\widehat\theta(\m)\right)
\nonumber\\
&=&
C_\m\left(\widehat\theta(\k)\widehat\theta^*(\el)\widehat\theta^*(\m)
+ \widehat\theta^*(\k)\widehat\theta(\el)\widehat\theta(\m)\right),
\end{eqnarray}
where we have used the identities $\el\times\m=\el\times\k=\k\times\m$ 
and suppressed the time variable. It is well known that both $E$ and $Z$ 
are conserved for each individual triad. This can be readily verified by 
the fact that the coupling coefficients in Eq.\ (\ref{triad}), $C_\k,~C_\el$ 
and $C_\m$, satisfy
$$C_\k + C_\el + C_\m = 0 = \frac{C_\k}{k^\alpha} +
\frac{C_\el}{\ell^\alpha} + \frac{C_\m}{m^\alpha}.$$
Furthermore, the transfer of $E$ and $Z$ is from the intermediate 
wave number to both the larger and smaller wave numbers or vice versa
(note the signs of the coupling coefficients). The former behavior 
appears to have been observed in numerical simulations of 2D turbulence 
without exception. 

We now analyze the coupling coefficients $C_\k,~C_\el$ and $C_\m$ in
detail. As crude estimates that hold in general, these can be bounded 
by (assuming $k<l<m$)
\begin{eqnarray}
\label{coeff1}
|C_\k| &=& 
\frac{|(m^\alpha-\ell^\alpha)\,\el\times\m|}{m^\alpha \ell^\alpha}
< k\ell^{1-\alpha}, \nonumber\\
|C_\el| &=&
\frac{|(k^\alpha-m^\alpha)\,\el\times\m|}{k^\alpha m^\alpha}
<\ell k^{1-\alpha}, \nonumber\\
|C_\m| &=&
\frac{|(\ell^\alpha-k^\alpha)\,\el\times\m|}{\ell^\alpha k^\alpha}
<\ell k^{1-\alpha}.
\end{eqnarray}
where we have used $|\el\times\m|=|\el\times\k|\le k\ell$. Similar 
estimates were obtained in \cite{TD06b} (for $\alpha=1,2$) and in
\cite{Constantin08b} (for $\alpha=1$). For $\alpha>2$,
local triads (i.e. $k\ltw\ell\ltw m$) at small scales are effectively 
``turned off'' because all $C_\k,~C_\el$ and $C_\m$ tend to zero in the 
limit $k\to\infty$. Furthermore, the convergence is as rapid as 
$k^{2-\alpha}$. An immediate interpretation of this observation is that 
local triads can be relatively ineffective in the direct transfer of 
$\langle\theta^2\rangle$ compared with their nonlocal counterparts 
(see below). At the critical value $\alpha=2$, $C_\k,~C_\el$ and $C_\m$ 
can remain order unity for local triads that satisfy 
$|\el\times\m|\approx k^2$ and 
$|m^\alpha-\ell^\alpha|\approx|k^\alpha-m^\alpha|\approx|
\ell^\alpha-k^\alpha|\approx k^\alpha$.
A majority of local triads satisfy both of these conditions. They
are neither ultra ``thin'' nor nearly isosceles and correspond to 
relatively sharp estimates in Eq.\ (\ref{coeff1}), which reduce to 
$|C_\k|\approx|C_\el|\approx|C_\m|\approx1$. This means that local
triads at small scales in the usual vorticity equation are moderately
active. They can play a significant role in the direct transfer. Finally, 
for $\alpha<2$, the interaction coefficients of these same triads 
diverge as $k\to\infty$. Their divergence can be seen to be as rapid
as $k^{2-\alpha}$. This result suggests that for this case, local 
triads can play an overwhelmingly dominant role in the direct transfer.

Next, we turn to nonlocal triads. These are ``thin'' triads with the
wave numbers $k,~\ell$ and $m$ satisfying $k\ll\ell\ltw m$. For this case, 
$C_\k,~C_\el$ and $C_\m$ can be estimated as follows.
\begin{eqnarray}
\label{Coeff}
|C_\k| &=& 
\frac{|(m^\alpha-\ell^\alpha)\,\el\times\m|}{m^\alpha \ell^\alpha}
\approx \frac{\alpha\,k^2}{\ell^\alpha}, \nonumber\\
|C_\el| &=& 
\frac{|(k^\alpha-m^\alpha)\,\el\times\m|}{k^\alpha m^\alpha}
\approx \ell k^{1-\alpha}, \nonumber\\
|C_\m| &=& 
\frac{|(\ell^\alpha-k^\alpha)\,\el\times\m|}{\ell^\alpha k^\alpha}
\approx \ell k^{1-\alpha}.
\end{eqnarray}
In the limit $\ell\to\infty$ (while $k<\infty$), $C_\k$ vanishes,
but both $C_\el$ and $C_\m$ ($C_\el\approx-C_\m$) diverge as rapidly as
$\ell$. This implies a vigorous exchange of generalized enstrophy 
between the two neighboring wave numbers $\ell$ and $m$, mediated by a 
virtually non-participating distant wave number $k$. This ultra local 
transfer by nonlocal interactions is virtually independent of $\alpha$ 
as the divergence of $C_\el$ and $C_\m$ is insensitive to $\alpha$. This 
result implies that local transfer by nonlocal interactions is an 
intrinsic characteristic of
this family of 2D turbulence models. Note, however, that this transfer 
can be significant only when the spectrum of the generalized enstrophy 
is not steeper than $k^{-1}$ \cite{T08}. In other words, the generalized
enstrophy needs to be physically present at small scales in order to 
facilitate such a transfer. This suggests that for $\alpha>2$ (recall 
that local triads are dynamically inactive), the generalized enstrophy 
spectra can plausibly scale as $k^{-1}$ because steeper spectra are unable 
to support a non-negligible direct transfer. This universal scaling has 
been suggested by Schorghofer \cite{Schorghofer00} and Wantanabe and 
Iwayama \cite{Wantanabe04}. Their justification is that $\theta$ can be 
considered as a passive scalar, a view in accord with the present analysis.

In passing, it is worth mentioning that the divergence of $C_\el$ and 
$C_\m$ in nonlocal triads is probably the reason for numerical instability 
in simulations of 2D turbulence with inadequate diffusion because local 
triads with coupling coefficients of order unity are evidently well 
behaved. Support for this claim can be derived from common observations 
that numerical divergences occur as soon as the modes in the vicinity of 
the truncation wave number are excited and well before they acquire any 
considerable amount of enstrophy. The same instability problem persists 
for $\alpha>2$ although the weak activities of local triads in this case 
may reduce the severity of the instability to a certain extent. 

\section{Concluding remarks}

We have introduced the notion of effective degree of nonlinearity to 
quantify the small-scale dynamics of a family of generalized models 
of two-dimensional turbulence governed by a broad class of nonlinear 
transport equations. Here, the active scalar $\theta=(-\Delta)^{\alpha/2}\psi$ 
($\alpha>0$) is advected by the 
incompressible flow $\u=(-\psi_y,\psi_x)$, where $\psi$ is the 
stream function. We have argued that although the advection term is 
quadratic, the effective degree of nonlinearity of the small-scale
dynamics is not always quadratic and depends on $\alpha$. It has been 
found that the active scalar gradient dynamics are virtually linear for 
$\alpha\ge2$ and become nonlinear for $\alpha<2$. Furthermore, the 
degree of nonlinearity increases as $\alpha$ is decreased from 2, 
becoming quadratic at $\alpha=1$ and exceeding quadratic 
nonlinearity for $\alpha<1$. It is conceivable that credible theories 
of the family's dynamics, particularly those involving small scales, 
need to account for the dependence on $\alpha$ of the effective degree 
of nonlinearlity. 

We have also found that local triads at small scales are highly active
for $\alpha<2$, moderately active for $\alpha=2$ and virtually inactive
for $\alpha>2$. On the other hand, nonlocal triads are characterized by 
a vigorous exchange of generalized enstrophy between pairs of neighboring 
wave numbers, mediated by the third non-participating distant wave number. 
This property is common for all $\alpha$, thereby implying that nonlocal 
interactions (but ultra local transfer) can be considered universal. In 
the absence of local triad activity ($\alpha>2$), this ultra local transfer 
is responsible for the direct transfer of generalized enstrophy. This
is similar to the problem of passive scalar transport by a large-scale 
flow as the weak feedback on the advecting flow by the active scalar 
can be neglected \cite{T08}. In this case, it appears plausible that 
generalized enstrophy spectra scale as $k^{-1}$. 

The local nature of the generalized enstrophy transfer can be seen to be 
unambiguous in the present study. In general, this transfer is local in 
wave number space regardless of what types of triads make the most 
contribution. For local triads the generalized enstrophy transfer is 
inherently local. For nonlocal interactions, the transfer is even 
``more'' local, having a relatively higher degree of locality compared 
to the transfer by local triads. More importantly, the transfer between 
distant wave numbers is largely insignificant. Hence, it makes sense to 
speak of the degree of locality of the direct generalized enstrophy 
transfer rather than to distinguish between local and distant transfer.

\end{document}